\documentclass[aps,prl,twocolumn]{revtex4}
\usepackage{amsmath}
\usepackage{graphicx}
\usepackage{dcolumn}
\usepackage{longtable}

\allowdisplaybreaks

\newcommand{\be}{\begin{equation}}
\newcommand{\ee}{\end{equation}}
\newcommand{\bea}{\begin{eqnarray}}
\newcommand{\eea}{\end{eqnarray}}
\newcommand{\bse}{\begin{subequations}}
\newcommand{\ese}{\end{subequations}}

\newcommand{\crr}{\nonumber \\}

\begin{document}

\title{Recoil corrections to the energy levels of hydrogenic atoms}

\author{Gregory S. Adkins}
\email[]{gadkins@fandm.edu}
\author{Jonathan Gomprecht}
\author{Yanxi Li}
\author{Evan Shinn}
\affiliation{Franklin \& Marshall College, Lancaster, Pennsylvania 17604}

\date{\today}

\begin{abstract}
We have completed the calculation of pure-recoil corrections of order $(Z \alpha)^6$ to Coulombic bound states of two spin-1/2 fermions without approximation in the particle masses.  Our result applies to systems of arbitrary mass ratio such as muonium and positronium, and also hydrogen and muonic hydrogen (with the neglect of proton structure effects).  We have shown how the two-loop master integrals that occur in the relativistic region can be computed in analytic form, and suggest that the same method can be applied to the three-loop integrals that would be present in a calculation of order $(Z \alpha)^7$ corrections.
\end{abstract}

\maketitle


Quantum field theory (QFT) describes the scattering of elementary particles in a relatively straightforward way; the description of bound states in QFT is less direct, but no less important.  The theory of bound states in QFT typically involves quantities that are of infinite order in the usual ``small'' parameters of the theory.  Despite the complexity, deep understanding of bound systems is required for the description of most objects that make up our world: protons and neutrons, nuclei, atoms, molecules, etc.  Two-body bound states of elementary particles such as positronium, muonium, and quarkonium allow for the cleanest description, uncluttered by any internal structure of the constituents.  Two-body systems bound by the Coulomb force are the most well understood since, compared to the strong force, the electromagnetic force lacks a confining phase and is weakly coupled on all scales of practical interest.  Consequently, the study of ``atoms'' such as positronium and muonium allows for the deep quantitative study of binding in QFT.  The two-body atom hydrogen is  of central interest for practical reasons despite the complication of having to take proton structure into account.  In fact, making a virtue out of necessity, high precision comparison of experiment and theory for hydrogen transition energies allows for the determination of the proton charge radius and other internal properties \cite{Tiesinga21}.  Some useful reviews of the theory of two-body bound states in quantum electrodynamics (QED) include \cite{Sapirstein90,Karshenboim05,Eides07,Yerokhin19}.

In this work we will focus on recoil corrections to the energy levels of two-body bound systems composed of elementary spin-1/2 fermions in their S states.  We label the masses of these particles $m_1$ and $m_2$, with $m_1$ typically the smaller of the two.  Examples of such systems include muonium, positronium, and hydrogen (although proton structure corrections mix with recoil corrections in an important way for hydrogen).  Recoil corrections in the nonrelativistic problem are completely accounted for by writing the Schr\"odinger-Coulomb equation in terms of the reduced mass $m_r=m_1 m_2/(m_1+m_2)$.  The Bohr energy level formula $-\frac{m_r (Z \alpha)^2}{2 n^2}$ takes recoil into account.  (Here $Z e$ is the charge of the positive constituent with $e$ the magnitude of the electron charge.  The fine structure constant is defined through $e^2= 4 \pi \alpha$.  We use units for which $\hbar = c = 1$.)  Relativistic corrections to the Bohr levels involve higher powers of $v^2/c^2$, and since $v \sim (Z \alpha) c$ for nonrelativistic Coulombic systems, the first corrections have relative order $(Z \alpha)^2$.  Breit \cite{Breit29} realized that these corrections could be found by considering a two-body hamiltonian consisting of free relativistic hamiltonians for each constituent plus a term describing one-photon exchange.  Fermi \cite{Fermi30} worked out the $(s=1)$ minus $(s=0)$ hyperfine splitting (hfs) contribution at this order, which comes entirely from one-photon exchange.  The Fermi term contains the square of the wave function at contact as a factor.  Breit and Meyerott \cite{Breit47} justified the use of the reduced mass in $\vert \psi_n(0) \vert^2 =  (m_r Z \alpha)^3/(\pi n^3)$ for the Fermi correction.   Explicit expectation values were worked out for the effective hamiltonian including first relativistic corrections plus one photon exchange by Barker and Glover \cite{Barker55}, with the result
\bea \label{DeltaE4}
\Delta E^{(4)} &=& \frac{m_r (Z \alpha)^4}{n^3} \bigg \{ -\frac{1}{2} + \frac{3}{8n} \crr
&\hbox{}& \hspace{0.4cm} + \frac{m_r^2}{m_1 m_2} \left( -\frac{1}{8n} + \frac{2}{3} \langle \vec \sigma_1 \cdot \vec \sigma_2 \rangle \right) \bigg \}
\eea
for the S-state energy correction at $O((Z \alpha)^4)$.  The spin operator has the expectation values $\langle \vec \sigma_1 \cdot \vec\sigma_2 \rangle_{s=1} = 1$ and $\langle \vec \sigma_1 \cdot \vec\sigma_2 \rangle_{s=0} = -3$, so that the hfs is $\langle \vec \sigma_1 \cdot \vec\sigma_2 \rangle_{\rm hfs} = 4$ and the spin average $(3(s=1)+(s=0))/4$ is $\langle \vec \sigma_1 \cdot \vec\sigma_2 \rangle_{\rm avg} = 0$.  The hfs for $n=1$ at order $(Z \alpha)^4$ defines the ``Fermi constant''
\be
E_F \equiv \frac{8 m_r^3 (Z \alpha)^4}{3 m_1 m_2} \, .
\ee
At order $(Z \alpha)^5$ the energy correction is more involved:
\bea \label{DeltaE5}
\Delta E^{(5)} &=& \frac{m_r^3 (Z \alpha)^5}{m_1 m_2 \pi n^3} \bigg \{ \frac{2}{3} \ln \left( \frac{1}{Z \alpha} \right) - \frac{8}{3} \ln k_0(n,0) \crr
&\hbox{}& \hspace{-0.6cm} + \frac{14}{3} \left ( H_n - \frac{1}{2n} + \ln \left( \frac{2}{n} \right) \right ) + \frac{41}{9} \crr
&\hbox{}& \hspace{-0.6cm} - 2 \frac{ m_2^2 \ln \left( m_1/m_r \right) - m_1^2 \ln \left( m_2/m_r \right) }{m_2^2-m_1^2} \crr
&\hbox{}& \hspace{-0.6cm} - \frac{2 m_1 m_2}{m_2^2-m_1^2} \ln \left( \frac{m_2}{m_1} \right) \langle \vec \sigma_1 \cdot \vec \sigma_2 \rangle \bigg \} \, .
\eea
The hyperfine contribution in $\Delta E^{(5)}$ \cite{Arnowitt53,Fulton54,Newcomb55} (proportional to $\langle \vec \sigma_1 \cdot \vec \sigma_2 \rangle$) comes entirely from two-photon exchange in the ``hard'' (relativistic) region of integration.  There are two relativistic scales corresponding to $m_1$ and $m_2$, and integration over the region between the two leads to the characteristic logarithmic dependence on the mass ratio.  The spin-independent contribution \cite{Salpeter52,Fulton54,Erickson65,Grotch69,Erickson77} is more complicated, having contributions from all three energy regions: hard (relativistic), soft (of order $m_r Z \alpha$), and ultrasoft (of order $m_r (Z \alpha)^2$).  The quantity $H_n$ is the $n^{\rm th}$ harmonic number $H_n = \sum_{i=1}^n \frac{1}{i}$ and $\ln k_0(n,\ell)$ is the Bethe log.  A table of Bethe logs can be found, for example, in \cite{Eides07}.  Detailed modern derivations of $\Delta E^{(5)}$ are given in Chs. 15 and 17 of \cite{Jentschura22}.

Work on order $(Z \alpha)^6$ recoil corrections to the hyperfine splitting commenced in 1971 \cite{Fulton71}. The logarithmic corrections were known by 1977 \cite{Lepage77,Bodwin78}, and the constant to go with the log (but not its complete mass dependence) was given by Bodwin, Yennie, and Gregorio \cite{Bodwin82}:
\be \label{DeltaE6hfs}
\Delta E^{(6)}_{\rm hfs} = E_F \frac{m_r^2 (Z \alpha)^2}{m_1 m_2 n^3} \left \{ 2 \ln \left( \frac{1}{Z \alpha} \right) - 8 \ln 2 + \frac{65}{18} \right \} \, .
\ee
 A review of this work is contained in \cite{Bodwin85}.  Later, Pachucki found the full state and mass dependence of the hfs recoil correction at order $(Z \alpha)^6$ \cite{Pachucki97}.  The mass dependence was obtained as the result of a numerical integration.  Work on recoil corrections to the spin-averaged energy shift (also referred to as the Lamb shift) at order $(Z \alpha)^6$ commenced in 1988 \cite{Erickson88}.  By 1993, after a number of false starts, it was clear that there was no $\ln(Z \alpha)$ contribution at this order \cite{Khriplovich93,Fell93}.  The complete correction at this order was given by Pachucki and Grotch in 1995 \cite{Pachucki95}.  Despite some controversy over this result \cite{Elkhovskii96,Eides97,Yelkhovsky98}, it was confirmed with complete state dependence by Eides and Grotch \cite{Eides97}:
 \bea \label{DeltaE6avg}
 \Delta E^{(6)}_{\rm avg} &=& \frac{m_1^2 (Z \alpha)^6}{m_2 n^3} \bigg \{ \frac{1}{8} + \frac{3}{8n} - \frac{1}{n^2} + \frac{1}{2n^3} \crr
 &\hbox{}& \hspace{0.6cm} + \left( 4 \ln 2 - \frac{7}{2} \right) \bigg \} \, .
 \eea
The result $\Delta E^{(6)}_{\rm avg}$ was further confirmed by a high-precision numerical evaluation of the order $m_1/m_2$ Lamb shift recoil correction \cite{Yerokhin15,Yerokhin16}.  Higher recoil corrections at order $(Z \alpha)^6$ were obtained by Blokland, Czarnecki, and Melnikov \cite{Blokland02} as a power series in $(m_1/m_2)$, with results up to order $(m_1/m_2)^4$.

Expressions (\ref{DeltaE4}) and (\ref{DeltaE5}) for $\Delta E^{(4)}$ and $\Delta E^{(5)}$ are ``pure recoil'' energy corrections.  They are proportional to a power of $Z \alpha$ with always the same number of interactions on the electron line as on the proton line and no radiative photons (emitted and absorbed on a single line) or vacuum polarization loops included.  They are exact functions of the particle masses.  Expressions (\ref{DeltaE6hfs}) and (\ref{DeltaE6avg}) for the order $(Z \alpha)^6$ contributions to the hfs and Lamb shift give recoil corrections but, except for the exact coefficient of the log term and Pachucki's numerical evaluation of the hfs \cite{Pachucki97}, these corrections are only known as a series expansion in the recoil parameter $m_1/m_2$.  In this work we obtain the S-state pure recoil correction at order $(Z \alpha)^6$ exact in the particle masses.  This completes, at last, the project of computing $O((Z \alpha)^6)$ pure recoil corrections for these states, which had previously only been obtained as an approximate function of the masses.

In this work we focus on S-state corrections.  Recoil corrections at order $(Z \alpha)^6$ for states with $\ell>0$ are discussed in Refs.~\cite{Khriplovich93,Fell93,Golosov95,Elkhovskii96,Zatorski22}.  The corrections for states with higher angular momentum do not involve the ``hard'' (relativistic) momenta that are a central challenge for the S-state corrections considered here.

\begin{figure}[t]
\includegraphics[width=3.28in]{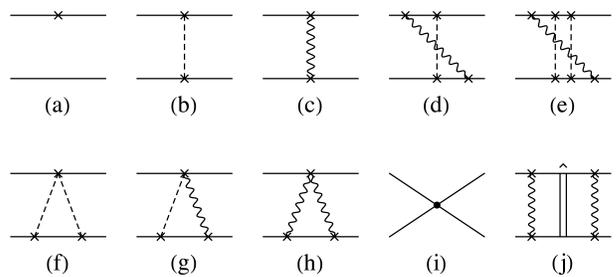}
\caption{\label{fig1} Kernels contributing recoil corrections at order $(Z \alpha)^6$.  The electron line is shown on the top and the positive particle (proton, positive muon, etc.) on the bottom.  The dotted line represents a Coulomb photon and the wiggly line a transverse photon.  The vertices are the full NRQED vertices for the interaction shown.  The kernels include (a) the relativistic kinetic energy correction, (b) Coulomb exchange with higher-order NRQED vertices, (c) transverse photon exchange with NRQED vertices, (d) crossed transverse-Coulomb photons, (e) crossed transverse-Coulomb-Coulomb photons, (f)-(h) ``$\Lambda$'' kernels with a single seagull vertex, (i) the contact term, and (j) the contribution from second-order perturbation theory.  Also included but not pictured are crossed graphs flipped left to right and seagull ``V'' graphs with the seagull vertex on the bottom instead of the top.}
\end{figure}

Our calculation was done using nonrelativistic QED (NRQED) \cite{Caswell86,Kinoshita96}.  Ultraviolet and infrared divergences were regulated using dimensional regularization \cite{Manohar97,Pineda98,Czarnecki99}.  The NRQED Feynman rules were read off of the Lagrangian given by Hill, Lee, Paz, and Solon \cite{Hill13} (see also \cite{Haidar20}).  Bound state energies were computed using the NRQED Bethe-Salpeter equation--the procedure is described in  \cite{Adkins18,Adkins19}.  Properties of the bound state wave function and expectation values in $D=3-2\epsilon$ spatial dimensions are given in \cite{Adkins18b,Adkins20}.  There are three main classes of recoil contributions at order $(Z \alpha)^6$: expectation values of the interaction kernels shown in Fig.~1(a)-(h), the expectation value of the contact kernel Fig.~1(i), and the second order perturbation contribution illustrated in Fig.~1(j).  The internal momenta of these NRQED expectation values are restricted to the non-relativistic region through use of the method of regions \cite{Beneke98,Jantzen11} (although the momenta involved in finding the ``matching'' coefficients \cite{Caswell86,Kinoshita96,Manohar97} for the contact term Fig.~1 (i) are hard).  The second-order perturbation contribution was computed as in \cite{Czarnecki99,Zatorski08}.  The totals of the two soft contributions (Fig.~1(a)-(h) and Fig.~1(j)) to the hfs are found to be
\bea
\Delta E^{\rm hfs}_{\rm soft} &=& \frac{\pi \vert \psi_n(0) \vert^2 (Z \alpha)^3 \bar \mu^{2 \epsilon}}{3 m_1 m_2} \bigg \{ \left ( \frac{44}{3} + \frac{12}{n} - \frac{44}{3n^2} \right ) \crr
&\hbox{}& \hspace{-0.6cm} + \frac{m_r^2}{m_1 m_2} \left ( \frac{4}{\tilde \epsilon} + \frac{62}{9} + \frac{24}{n} + \frac{32}{3n^2} \right ) \bigg \} \, ,
\eea
where
\be
\frac{1}{\tilde \epsilon} = \frac{1}{\epsilon} + 4 \ln \left( \frac{\mu n}{2 m_r Z \alpha} \right) - 4 H_n \, .
\ee
(Here $\mu$ is the mass parameter introduced in the process of dimensional regularization, and $\bar \mu^2=\mu^2 e^{\gamma_E}/(4\pi)$ with $\gamma_E \approx 0.57722$ the Euler-Mascheroni constant.  The product of the charges is $q_1 q_2 = - 4 \pi \alpha \bar \mu^{2\epsilon}$.)  The probability density of contact in $D$ dimensional space is
\be
\vert \psi_n(0) \vert^2 = \frac{(m_r Z \alpha)^3}{\pi n^3} + O(\epsilon) \, .
\ee
For the soft contribution to the spin-averaged energy correction we find the finite result
\bea
\Delta E^{\rm avg}_{\rm soft} &=& \frac{m_r (Z \alpha)^6}{n^3} \bigg \{ \left ( -\frac{1}{8} - \frac{3}{8n} + \frac{3}{4n^2} - \frac{5}{16n^3} \right ) \crr
&\hbox{}& \hspace{-0.6cm} + \frac{m_r^2}{m_1 m_2} \left ( -\frac{1}{4n^2} + \frac{3}{16n^3} \right ) \crr
&\hbox{}& \hspace{-0.6cm} + \frac{m_r^4}{(m_1 m_2)^2} \left ( -\frac{4}{9} - \frac{2}{n} - \frac{1}{3n^2} - \frac{1}{16n^3} \right ) \bigg \} \, .
\eea

\begin{figure}[t]
\includegraphics[width=2.8in]{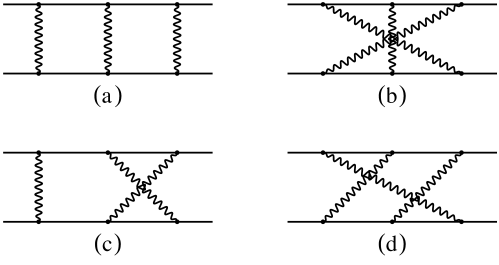}
\caption{\label{fig2} Three-photon-exchange scattering diagrams contributing to the contact-term matching coefficient of Fig.~1(i).  The contributions of (c) and (d) must be doubled to account for equal reflected diagrams.}
\end{figure}

The energy correction coming from the contact term Fig.~1 (i) can be expressed in terms of the two-particle threshold (zero relative velocity) scattering amplitudes ${\cal M}^s$ of Fig.~2 by
\be 
\Delta E_{\rm hard}^s = - \vert \psi(0) \vert^2 {\cal M}^s \, .
\ee
These scattering amplitudes contain momenta exclusively from the hard, or relativistic region.  We calculate the amplitudes using standard QED in Feynman gauge.  The $D$-dimensional traces were done using FeynCalc \cite{Mertig91,Shtabovenko16}.  Then the ``integration by parts''  identities were implemented through use of the program FIRE \cite{Smirnov20}.  Some of the master integrals obtained from FIRE were integrable using standard techniques \cite{Smirnov12}.  The rest were integrated using the method of differential equations \cite{Kotikov91,Bern93,Gehrmann00} in terms of the variable $x=m_1/m_2$.  The (first order, coupled) differential equations were put into a canonical form \cite{Henn15} using Fuchsia \cite{Gituliar17}, and the solutions were expressed in terms of harmonic polylogarithms $\text{HPL}(\{a \},x)$ \cite{Remiddi00,Maitre06}.  The master integrals were expanded (using the method of regions) about the point $x=0$ (as in \cite{Blokland02}) since the integrals with $m_1 \rightarrow 0$ are tractable.  The $x=0$ limits of the integrals were used as boundary conditions, which along with the differential equations allowed us to solve for the master integrals for all positive values of $x$.  We found that the amplitudes can be expressed as

\be
{\cal M}^s = - \frac{ \pi (Z \alpha)^3}{m_1 m_2} \bar \mu^{2 \epsilon} \left ( \frac{\mu^2}{m_r^2} \right )^{2 \epsilon} {\cal H}^s
\ee
with 
\bse \bea
{\cal H}_{\rm hfs} &=& {\cal H}^{s=1}-{\cal H}^{s=0} = - \frac{4 x}{3 (1+x)^2 \epsilon} + h_{\rm hfs} \, , \\[3pt]
{\cal H}_{\rm avg} &=& \frac{D {\cal H}^{s=1} + {\cal H}^{s=0}}{D+1} = h_{\rm avg} \, .
\eea \ese

The hfs and spin-average functions are
\bea
h_{\rm hfs}(x) &=&
\frac{1}{9 \pi^2 (1-x^2)^2} \Big \{  2 \pi ^2 x \left(x^2-29\right) (x-1) \crr
&\hbox{}& \hspace{-1.4cm} + 144 \pi ^2 x \left(x^2-1\right) \log (2) - 72 x^2 \left(3 x^2-11\right) \zeta (3) \crr
&\hbox{}& \hspace{-1.4cm}  + 8 \pi ^2 x^2 \left(x^2+3\right) \log (x) + 12 x^2 \left(x^2-3\right) \log^2(x) \crr
&\hbox{}& \hspace{-1.4cm}  + 12 \pi ^2 (x+1)^3 (x-1) \text{HPL}(\{1\},x) \crr
&\hbox{}& \hspace{-1.4cm}  + 12 \pi ^2 \left(x-1 \right)^2 \left(x^2-3\right) \text{HPL}(\{-1\},x) \crr
&\hbox{}& \hspace{-1.4cm} + 12 \left(x^2-1\right) \left(x^2-10 x+1\right)\text{HPL}(\{1,0\},x) \crr
&\hbox{}& \hspace{-1.4cm} - 12 \left(x^2-1\right) \left(x^2+10 x+1\right)\text{HPL}(\{-1,0\},x) \crr
&\hbox{}& \hspace{-1.4cm}  - 48 (x+1)^2 \left(x^2+2 x-2\right) \text{HPL}(\{1,0,0\},x) \crr
&\hbox{}& \hspace{-1.4cm}  + 48 (x-1)^2 \left(x^2-2 x-2\right) \text{HPL}(\{-1,0,0\},x) \crr
&\hbox{}& \hspace{-1.4cm}  - 48 (x+1) \left(2 x^3-6 x^2+3 x-1\right)\text{HPL}(\{2,0\},x) \crr
&\hbox{}& \hspace{-1.4cm}  + 48 (x-1) \left(2 x^3+6 x^2+3 x+1\right) \text{HPL}(\{-2,0\},x) \crr
&\hbox{}& \hspace{-1.4cm} + 144 (x+1) (x-1)^3 \text{HPL}(\{-1,1,0\},x) \crr
&\hbox{}& \hspace{-1.4cm} + 144 (x-1) (x+1)^3 \text{HPL}(\{1,-1,0\},x) \Big \} \, ,
\eea
and
\bea
h_{\rm avg}(x) &=& \frac{1}{18 \pi^2 x (1-x^2)^2} \Big \{ - 72 \pi ^2 x \left(x^2-1\right) \log (2) \crr
&\hbox{}& \hspace{-1.4cm} + 3 \pi ^2 x (x-1) \left(3 x^3+19 x^2-5 x+9\right) \crr
&\hbox{}& \hspace{-1.4cm}  - 108 x^2 \left(x^2-2\right) \left(x^2-1\right) \zeta (3) \crr
&\hbox{}& \hspace{-1.4cm}  + 12 x^2 \left( x^2-1 \right) \left [ \pi^2 \left(x^2 - 4 \right) -3 \right ] \log (x) 
+ 36 x^4 \log ^2(x) \crr
&\hbox{}& \hspace{-1.4cm}  + 6 \pi ^2 (x-1) (x+1)^3 \left(x^2-3 x+1\right)\text{HPL}(\{1\},x) \crr
&\hbox{}& \hspace{-1.4cm}  - 6 \pi ^2 (x-1)^2 (x+1) \left(x^3-x^2+7 x+5\right)\text{HPL}(\{-1\},x) \crr
&\hbox{}& \hspace{-1.4cm}  + 36 x \left(x^2-1\right) \left(2 x^2+x+2\right) \text{HPL}(\{1,0\},x)\crr
&\hbox{}& \hspace{-1.4cm}  + 36 x \left(x^2-1\right)\left(2 x^2-x+2\right) \text{HPL}(\{-1,0\},x) \crr
&\hbox{}& \hspace{-1.4cm}  + 36 (x-1) (x+1)^2 \left(x^2+3 x-2\right)\text{HPL}(\{1,0,0\},x)  \crr
&\hbox{}& \hspace{-1.4cm}  + 36 (x+1) (x-1)^2 \left(x^2-3 x-2\right) \text{HPL}(\{-1,0,0\},x) \crr
&\hbox{}& \hspace{-1.4cm}  - 36 x (x+1) (x-1)^2 \left(2 x^2+3 x-1\right) \text{HPL}(\{2,0\},x) \crr
&\hbox{}& \hspace{-1.4cm}  + 36 x (x-1) (x+1)^2 \left(2 x^2-3 x-1\right) \text{HPL}(\{-2,0\},x) \crr
&\hbox{}& \hspace{-1.4cm} + 72 (x+1) (x-1)^3 \left(x^2+3 x+1\right) \text{HPL}(\{-1,1,0\},x) \crr
&\hbox{}& \hspace{-1.4cm} + 72 (x-1) (x+1)^3\left(x^2-3 x+1\right) \text{HPL}(\{1,-1,0\},x) \Big \} \, . \crr
\eea
A non-trivial consistency check of our results for the $h(x)$ functions comes from the symmetry under interchange of particle masses $m_1 \leftrightarrow m_2$, which implies $h(1/x) = h(x)$.

The final results (adding the soft plus hard contributions) are
\bea
\Delta E_{\rm hfs} &=& \frac{8 m_r^3 (Z \alpha)^6}{3 m_1 m_2 n^3} \bigg \{ \left( \frac{11}{6} + \frac{3}{2n} - \frac{11}{6n^2} + \frac{3}{8} h_{\rm hfs}(x) \right) \crr
&\hbox{}& \hspace{-1.6cm} + \frac{m_r^2}{m_1 m_2} \left ( 2 \ln \left( \frac{n}{2 Z \alpha} \right) - 2 {\rm H}_n+ \frac{31}{36} + \frac{3}{n} + \frac{4}{3n^2} \right ) \bigg \} 
\eea
since $x/(1+x)^2 = m_r^2/(m_1 m_2)$, and
\bea
\Delta E_{\rm avg} &=& \frac{m_r (Z \alpha)^6}{n^3} \bigg \{ \left( - \frac{1}{8} - \frac{3}{8n} + \frac{3}{4n^2} - \frac{5}{16n^3} \right) \crr
&+& \frac{m_r^2}{m_1 m_2} \left ( - \frac{1}{4n^2} + \frac{3}{16n^3} + h_{\rm avg}(x)\right ) \crr
&+& \frac{m_r^4}{m_1^2 m_2^2} \left( - \frac{4}{9} - \frac{2}{n} - \frac{1}{3n^2} - \frac{1}{16n^3} \right) \bigg \} \, .
\eea
For the two spin states separately we have
\bse \bea
\Delta E_{s=1} &=& \Delta E_{\rm avg} + \frac{1}{4} \Delta E_{\rm hfs} \, , \\[3pt]
\Delta E_{s=0} &=& \Delta E_{\rm avg} - \frac{3}{4} \Delta E_{\rm hfs} \, .
\eea \ese

Our results are in complete agreement with Pachucki's numerical evaluation of the mass dependence of the hfs contribution \cite{Pachucki97}.  In particular, we reproduce his graph of Fig.~3.  It is also easy to expand our results for small values of $x=m_1/m_2$.  We find a few errors in the series expansions of Blokland {\it et al.} \cite{Blokland02}.  In their hfs result of Eq. (22) in the order $(m/M)$ term the $-\frac{13}{12}$ should be $+\frac{11}{12}$, and in the order $(m/M)^2$ term of that equation the $-13 \ln 2$ should be $-12 \ln 2$ and $-\frac{4}{3}$ should be $-\frac{7}{3}$.  In their average result of Eq. (23) the $-\frac{113}{18}$ should be $-\frac{133}{18}$.

For particle-antiparticle bound systems such as positronium, the $x \rightarrow 1$ limit is required.  This limit can be easily obtained using the forms of $h(x)$ given above (after using the HPL command ``HPLConvertToKnownFunctions'').  The required limits are
\bse \bea
h_{\rm hfs}(1) &=& \frac{1}{\pi^2} \left \{ -17 \zeta(3) - \frac{2}{3} \pi^2 \ln 2 + \frac{10}{3} \right \} \, , \\[3pt]
h_{\rm avg}(1) &=& \frac{1}{\pi^2} \left \{ -3 \zeta(3) - \frac{13 \pi^2}{24} - \frac{11}{2} \right \} \, .
\eea \ese
For particle-antiparticle bound systems the ``recoil'' corrections to the energies at $O(\alpha^6)$ are
\bse \bea
\Delta E_{\rm hfs} &=& \frac{m \alpha^6}{n^3} \bigg \{\frac{1}{6} \ln \left( \frac{1}{\alpha} \right) - \frac{17 \zeta(3)}{8\pi^2} + \frac{5}{12\pi^2} - \frac{\ln 2}{4} \crr
&\hbox{}& \hspace{0.0cm} + \frac{295}{432} + \frac{1}{6} \ln n - \frac{1}{6} {\rm H}_n + \frac{3}{4n} - \frac{1}{2n^2} \bigg \} \, , \\[3pt]
\Delta E_{\rm avg} &=& \frac{m \alpha^6}{n^3} \bigg \{ -\frac{3 \zeta(3)}{8\pi^2} - \frac{11}{16\pi^2} - \frac{83}{576} \crr
&\hbox{}& \hspace{0.6cm} - \frac{1}{4n} + \frac{1}{3n^2} - \frac{69}{512n^3} \bigg \} \, .
\eea \ese
We have used $Z=1$, $m_1 = m_2 = m$, $m_r=m/2$, $x=m_1/m_2=1$ for this evaluation.  The results here are in agreement with earlier evaluations: Refs.~\cite{Pachucki97,Czarnecki99,Adkins98,Burichenko01} for the hyperfine correction and Ref.~\cite{Czarnecki99} for the average energy shift.

In summary, we have completed the calculation of pure recoil corrections at order $\alpha^6$ to two-fermion Coulombic bound systems.  One reason for performing this calculation was simply to get the exact result that would apply to diverse systems such as hydrogen, muonium, muonic hydrogen, and positronium without the need for  approximations or expansions.  The second important reason was to test a method for calculating the extremely difficult master integrals that will be required for the order $\alpha^7$ hard corrections to two-body Coulombic bound states.  Even at order $\alpha^6$, some of the master integrals are challenging to handle by traditional methods (see \cite{Broadhurst90} for example), and the three-loop massive integrals required at order $\alpha^7$ are anticipated to be significantly more difficult.


\begin{acknowledgments} 
This material is based upon work supported by the National Science Foundation under Grant No. PHY-2011762 and by Franklin \& Marshall College through the Hackman Scholars Program.
\end{acknowledgments}





\begin{thebibliography}{10}

\bibitem{Tiesinga21} E. Tiesinga, P. J. Mohr, D. B. Newell, and B. N. Taylor, CODATA recommended values of the fundamental physical constants: 2018, Rev. Mod. Phys. {\bf 93}, 025010 (2021).
\bibitem{Sapirstein90} J. R. Sapirstein and D. R. Yennie, Theory of hydrogenic bound states, in ``Quantum Electrodynamics'', ed. by. T. Kinoshita (World Scientific, Singapore, 1990), p. 560.
\bibitem{Karshenboim05} S. G. Karshenboim, Precision physics of simple atoms: QED tests, nuclear structure and fundamental constants, Phys. Rep. {\bf 422}, 1 (2005).
\bibitem{Eides07} M. I. Eides, H. Grotch, and V. A. Shelyuto, Theory of light hydrogenic bound states, (Springer, Berlin, 2007).
\bibitem{Yerokhin19} V. A. Yerokhin, K. Pachucki, and V. Patk\'o\v{s}, Theory of the Lamb shift in hydrogen and light hydrogen-like ions, Annalen der Physik {\bf 531}, 1800324 (2019).
\bibitem{Breit29} G. Breit, The effect of retardation on the interaction of two electrons, Phys. Rev. {\bf 34}, 553 (1929).
\bibitem{Fermi30} E. Fermi, \"Uber die magnetischen momente der atomkerne, Z. Phys. {\bf 60}, 320 (1930).
\bibitem{Breit47} G. Breit and R. E. Meyerott, Effect of nuclear motion of the hyperfine structure of the ground term of hydrogen, Phys. Rev. {\bf 72}, 1023 (1947).
\bibitem{Barker55} W. A. Barker and F. N. Glover, Reduction of relativistic two-particle wave equations to approximate forms. III, Phys. Rev. {\bf 99}, 317 (1955).
\bibitem{Arnowitt53} R. Arnowitt, The hyperfine structure of hydrogen, Phys. Rev. {\bf 92}, 1002 (1953).
\bibitem{Fulton54} T. Fulton and P. C. Martin, Two-body system in quantum electrodynamics.  Energy levels of positronium, Phys. Rev. {\bf 95}, 811 (1954).
\bibitem{Newcomb55} W. A. Newcomb and E. E. Salpeter, Mass corrections to the hyperfine structure in hydrogen, Phys. Rev. {\bf 97}, 1146 (1955).
\bibitem{Salpeter52} E. E. Salpeter, Mass corrections to the fine structure of hydrogen-like atoms, Phys. Rev. {\bf 87}, 328 (1952).
\bibitem{Erickson65} G. W. Erickson and D. R. Yennie, Radiative level shifts, I. Formulation and lowest order Lamb shift, Ann. Phys. (N.Y.) {\bf 35}, 271 (1965).
\bibitem{Grotch69} H. Grotch and D. R. Yennie, Effective potential model for calculating nuclear corrections to the energy levels of hydrogen, Rev. Mod. Phys. {\bf 41}, 350 (1969).
\bibitem{Erickson77} G. W. Erickson, Energy levels of one-electron atoms, J. Phys. Chem. Ref. Data {\bf 6}, 831 (1977).
\bibitem{Jentschura22} U. D. Jentschura and G. S. Adkins, Quantum Electrodynamics: Atoms, Lasers and Gravity (World Scientific, Singapore, 2022).
\bibitem{Fulton71} T. Fulton, D. A. Owen, and W. W. Repko, Order $(m_e/m_\mu) \alpha^2 \ln \alpha^{-1}$ corrections to the muonium hyperfine structure, Phys. Rev. Lett. {\bf 26}, 61 (1971).
\bibitem{Lepage77} G. P. Lepage, Analytic bound-state solutions in a relativistic two-body formalism with applications in muonium and positronium, Phys. Rev. A {\bf 16}, 863 (1977).
\bibitem{Bodwin78} G. T. Bodwin and D. R. Yennie, Hyperfine splitting in positronium and muonium, Phys. Rep. {\bf 43}, 267 (1978).
\bibitem{Bodwin82} G. T. Bodwin, D. R. Yennie, and M. A. Gregorio, Corrections to the muonium hyperfine splitting of relative order $\alpha^2 (m_e/m_\mu)$, Phys. Rev. Lett. {\bf 48}, 1799 (1982).
\bibitem{Bodwin85} G. T. Bodwin, D. R. Yennie, and M. A. Gregorio, Recoil effects in the hyperfine structure of QED bound states, Rev. Mod. Phys. {\bf 57}, 723 (1985).
\bibitem{Pachucki97} K. Pachucki, Effective Hamiltonian approach to the bound state: Positronium hyperfine structure, Phys. Rev. A {\bf 56}, 297 (1997).
\bibitem{Erickson88} G. W. Erickson and H. Grotch, Lamb-shift recoil effects in hydrogen, Phys. Rev. Lett. {\bf 60}, 2611 (1988).  [Erratum: Phys. Rev. Lett. {\bf 63}, 1326 (1989).]
\bibitem{Khriplovich93} I. B. Khriplovich, A. I. Milstein, and A. S. Yelkhovsky, Logarithmic corrections in the two-body QED problem, Physica Scripta {\bf T46}, 252 (1993).
\bibitem{Fell93} R. N. Fell, I. B. Khriplovich, A. I. Milstein, and A. S. Yelkhovsky, On the recoil corrections in hydrogen, Phys. Lett. A {\bf 181}, 172 (1993).
\bibitem{Pachucki95} K. Pachucki and H. Grotch, Pure recoil corrections to hydrogen energy levels, Phys. Rev. A {\bf 51}, 1854 (1995).
\bibitem{Elkhovskii96} A. S. Elkhovski\u \i, Order $(Z \alpha)^4 (m/M) R_\infty$ correction to the hydrogen energy levels, JETP {\bf 83}, 230 (1996).  [Zh. \' Eksp. Teor. Fiz. {\bf 110}, 431 (1996).]
\bibitem{Eides97} M. I. Eides and H. Grotch, Recoil corrections of order $(Z \alpha)^6 (m/M) m$ to the hydrogen energy levels recalculated, Phys. Rev. A {\bf 55}, 3351 (1997).
\bibitem{Yelkhovsky98} A. S. Yelkhovsky, Recoil correction to hydrogen energy levels: A revision, J. Expt. Theor. Phys. {\bf 86}, 472 (1998).
\bibitem{Yerokhin15} V. A. Yerokhin and V. M. Shabaev, Nuclear recoil effect in the Lamb shift of light hydrogenlike atoms, Phys. Rev. Lett. {\bf 115}, 233002 (2015).
\bibitem{Yerokhin16} V. A. Yerokhin and V. M. Shabaev, Nuclear recoil corrections to the Lamb shift of hydrogen and light hydrogenlike ions, Phys. Rev. A {\bf 93}, 062514 (2016). 
\bibitem{Blokland02} I. Blokland, A. Czarnecki, and K. Melnikov, Expansion of bound-state energies in powers of $m/M$ and $(1-m/M)$, Phys. Rev. D {\bf 65}, 073015 (2002).
\bibitem{Golosov95} \'E. A. Golosov, A. S. Elkhovski\u \i, A. I. Mil'shte\u \i n, and I. B. Khriplovich, Order $\alpha^4 (m/M) R_\infty$ corrections to hydrogen P levels, JETP {\bf 80}, 208 (1995).  [Zh. \' Eksp. Teor. Fiz. {\bf 107}, 393 (1995).]
\bibitem{Zatorski22} J. Zatorski, V. Patk\'o\v{s}, and K. Pachucki, Quantum electrodynamics of two-body systems with arbitrary masses, arXiv:2207.14155 [physics.atom-ph].
\bibitem{Caswell86} W. E. Caswell and G. P. Lepage, Effective Lagrangians for bound state problems in QED, QCD, and other field theories, Phys. Lett. B {\bf 167}, 437 (1986).
\bibitem{Kinoshita96} T. Kinoshita and M. Nio, Radiative corrections to the muonium hyperfine structure: The $\alpha^2 (Z \alpha)$ correction, Phys. Rev. D {\bf 53}, 4909 (1996).
\bibitem{Manohar97} A. V. Manohar, Heavy quark effective theory and nonrelativistic QCD Lagrangian to order $\alpha_{\rm S}/m^3$, Phys. Rev. D {\bf 56}, 230 (1997).
\bibitem{Pineda98} A. Pineda and J. Soto, The Lamb shift in dimensional regularization, Phys. Lett. B {\bf 420}, 391 (1998).
\bibitem{Czarnecki99} A. Czarnecki, K. Melnikov, and A. Yelkhovsky, Positronium S-state spectrum: Analytic results at $O(m \alpha^6)$, Phys. Rev. A {\bf 59}, 4316 (1999).
\bibitem{Hill13} R. J. Hill, G. Lee, G. Paz, and M. P. Solon, NRQED Lagrangian at order $1/M^4$, Phys. Rev. D {\bf 87}, 053017 (2013).
\bibitem{Haidar20} M. Haidar, Z.-X. Zhong, V. I. Korobov, and J.-Ph. Karr, Nonrelativistic QED approach to the fine- and hyperfine-structure corrections of order $m \alpha^6$ and $m \alpha^6 (m/M)$: Application to the hydrogen atom, Phys. Rev. A {\bf 101}, 022501 (2020).
\bibitem{Adkins18} G. S. Adkins, Higher order corrections to positronium energy levels, J. Phys.: Conf. Ser. {\bf 1138}, 012005 (2018).
\bibitem{Adkins19} G. S. Adkins, B. Akers, M. F. Alam, L. M. Tran, and X. Zhang, Calculation of higher order corrections to positronium energy levels, Proc. Sci. (FFK2019) 004 (2019).
\bibitem{Adkins18b} G. S. Adkins, The hydrogen atom in $D=3-2 \epsilon$ dimensions, Phys. Lett. A {\bf 382}, 1545 (2018).
\bibitem{Adkins20} G. S. Adkins, M. F. Alam, C. Larison, and R. Sun, Coulomb expectation values in $D=3$ and $D=3-2\epsilon$ dimensions, Phys. Rev. A {\bf 101}, 042511 (2020).

\bibitem{Beneke98} M. Beneke and V. A. Smirnov, Asymptotic expansion of Feynman integrals near threshold, Nucl. Phys. B {\bf 522}, 321 (1998).
\bibitem{Jantzen11} B. Jantzen, Foundation and generalization of the expansion by regions, J. High Energy Phys. {\bf 2011}, 76 (2011).

\bibitem{Zatorski08} J. Zatorski, $O(m \alpha^6)$ corrections to energy levels of positronium with nonvanishing orbital angular momentum, Phys. Rev. A {\bf 78}, 032103 (2008).
\bibitem{Mertig91} R. Mertig, M. B\"ohm, and A. Denner, FeynCalc - computer-algebraic calculation of Feynman amplitudes, Comp. Phys. Comm. {\bf 64}, 345 (1991).
\bibitem{Shtabovenko16} V. Shtabovenko, R. Mertig, and F. Orellana, New developments in FeynCalc 9.0, Comp. Phys. Comm. {\bf 207}, 432 (2016).
\bibitem{Smirnov20} A. V. Smirnov and F. S. Chukharev, FIRE6: Feynman Integral REduction with modular arithmetic, Comp. Phys. Comm. {\bf 247}, 106877 (2020).
\bibitem{Smirnov12} V. A. Smirnov, Analytic Tools for Feynman Integrals (Springer, Heidelberg, 2012).
\bibitem{Kotikov91} A. V. Kotikov, Differential equations method. New technique for massive Feynman diagram calculation, Phys. Lett. B {\bf 254}, 158 (1991).
\bibitem{Bern93} Z. Bern, L. Dixon, and D. A. Kosower, Dimensionally regulated one-loop integrals, Phys. Lett. B {\bf 302}, 299 (1993).  [Erratum: Phys. Lett. B 318, 649 (1993).]
\bibitem{Gehrmann00} T. Gehrmann and E. Remiddi, Differential equations for two-loop four-point functions, Nucl. Phys. B {\bf 580}, 485 (2000).
\bibitem{Henn15} J. M. Henn, Lectures on differential equations for Feynman integrals, J. Phys. A {\bf 48}, 153001 (2015).
\bibitem{Gituliar17} O. Gituliar and V. Magerya, Fuchsia: A tool for reducing differential equations for Feynman master integrals to epsilon form, Comp. Phys. Comm. {\bf 219}, 329 (2017).
\bibitem{Remiddi00} E. Remiddi and J. A. M. Vermaseren, Harmonic polylogarithms, Int. J. Mod. Phys. A {\bf 15}, 725 (2000).
\bibitem{Maitre06} D. Ma\^{\i}tre, HPL, a Mathematica implementation of the harmonic polylogarithms, Comp. Phys. Comm. {\bf 174}, 222 (2006).
\bibitem{Adkins98} G. S. Adkins and J. Sapirstein, Order $m \alpha^6$ contributions to ground-state hyperfine splitting in positronium, Phys. Rev. A {\bf 58}, 3552 (1998).  [Erratum: Phys. Rev. A {\bf 61}, 069902(E) (2000).]
\bibitem{Burichenko01} A. P. Burichenko, ``Recoil''-effect-induced contribution of order $m \alpha^6$ to the hyperfine splitting of the positronium ground state, Phys. At. Nucl. {\bf 64}, 1628 (2001) [Yad. Fiz. {\bf 64}, 1709 (2001).]
\bibitem{Broadhurst90} D. J. Broadhurst, The master two-loop diagram with masses, Z. Phys. C {\bf 47}, 115 (1990).



\end{thebibliography}
\end{document}